\documentclass{revtex4}
\usepackage{graphicx}
\usepackage{fancyhdr}
\usepackage{amsmath}
\newcommand{\al}[1]{\begin{align}#1\end{align}}

\newcommand{\bb}{\begin{bmatrix}}
\newcommand{\eb}{\end{bmatrix}}

\usepackage{color}

\begin{document}

\title{Phenomenological constraints on universal extra dimensions \\ at LHC and electroweak precision test
}

%

\author{Takuya Kakuda$^1$, Kenji Nishiwaki$^2$, Kin-ya Oda$^3$, Naoya Okuda$^3$, Ryoutaro Watanabe$^4$}
\affiliation{1-Graduate School of Science and Technology, Niigata University, Niigata 950-2121, JAPAN\\
2-Regional Centre for Accelerator-based Particle Physics, Harich-Chandra Reseach Institute, Allahabad 211 019, INDIA\\
3-Department of Physics, Osaka University, Osaka 560-0043, JAPAN\\
4-Theory Group, KEK, Tsukuba, Ibaraki 305-0801, JAPAN
}

\begin{abstract}\noindent
We show two types of bounds on five- and six-dimensional universal extra dimension (UED) models from the latest results of the Higgs search at the LHC and of the electroweak precision data for the $S$ and $T$ parameters. 
The UED models on which we put lower bounds are the minimal UED model in five dimensions and the six dimensional ones on $T^2/Z_2$, $T^2/(Z_2 \times Z_2')$ , $ T^2/Z_4$, $S^2$, $S^2/Z_2$, $RP^2$ and projective sphere. 
The highest possible ultraviolet cutoff scale for each UED model is evaluated from the vacuum stability of the Higgs potential by solving the renormalization group equation of the Higgs self coupling.
The bounds on the KK scale in the minimal UED model is 650 GeV from the LHC results and  700 GeV from the $S, T$ analysis at the 95\% confidence level, 
while those in the several 6D UED models are $850\,\text{GeV} \sim 1350\,\text{GeV}$ (Higgs search) and $900\,\text{GeV} \sim 1500\,\text{GeV}$ ($S,T$ analysis).
\end{abstract}


\maketitle



\section{Introduction}
Even after the discovery of the Higgs-like particle at the Large Hadron Collider (LHC), phenomenology of the Higgs sector is not fully revealed. 
The ATLAS and CMS experiments reported their recent results on signal strengths of the Higgs like boson (defined as the ratio of Higgs signal cross sections between the experimental result and the SM prediction)
for its decay into diphoton ($\gamma\gamma$) and diboson ($ZZ$ and $WW$)~\cite{ATLAS:2013gamma,ATLAS:2013Z,ATLAS:2013W,CMS:2013gamma,CMS:2013Z,CMS:2013W}. 
According to these results, the signal strengths of $H \to \gamma\gamma$, $ZZ$ and $WW$ turn out to be $1.65\pm0.24^{+0.25}_{-0.18}$, $1.7^{+0.5}_{-0.4}$ and $1.01\pm0.31$ at the ATLAS experiment, 
while $0.78\pm0.27$ (MVA based), $0.91^{+0.30}_{-0.24}$ and $0.71\pm0.37$ (cut based) at the CMS experiment. 
These results are consistent with the SM but there still is a room for a new physics effect in these processes. 
In this work, we put bounds on universal extra dimension (UED) models.

The UED is a candidate of new physics, in which all the SM particles propagate in extra compactified spacial dimensions. 
The five-dimensional minimal UED (mUED) model without tree-level brane-localized term as a minimal extension of the SM, which is constructed on $S^1/Z_2$~\cite{Appelquist:2000nn}, has been well studied.
Six-dimensional UED models with various two-dimensional compactified spaces are also considered. 
We investigate the 6D UED models based on two torus, $T^2/Z_2$~\cite{Appelquist:2000nn}, $T^2/Z_4$~\cite{Dobrescu:2004zi,Burdman:2005sr}, $T^2/(Z_2 \times Z'_2)$~\cite{Mohapatra:2002ug}, on two sphere $S^2/Z_2$~\cite{Maru:2009wu} and $S^2$, and on the non-orientable manifolds, namely the real projective plane $RP^2$~\cite{Cacciapaglia:2009pa} and the projective sphere (PS)~\cite{Dohi:2010vc}, 
by putting bounds on the Kaluza-Klein (KK) scale from the results of the Higgs signal search and the electroweak precision measurements~\footnote{See also Ref.~\cite{Antoniadis:1990ew} for an earlier proposal of a TeV scale extra dimension.}. 
For details of these models, see for example Refs.~\cite{Nishiwaki:2011gm,Nishiwaki:2011gk}.

For bounds on the UED models from the electroweak precision measurements, we use the $S$ and $T$ parameters defined as in Ref.~\cite{Peskin:1990zt,Peskin:1991sw}, 
which are the quantities based on two point functions of the gauge bosons in the electroweak sector. 
The recent constraints on the $S$ and $T$ parameters are given in Ref.~\cite{Baak:2012kk}. 
For the bounds from the Higgs signal search, we use the recent results obtained in Ref.~\cite{ATLAS:2013gamma,ATLAS:2013Z,ATLAS:2013W,CMS:2013gamma,CMS:2013Z,CMS:2013W} for each decay process. 
In order to calculate these quantities in the UED models, we need to know an ultraviolet (UV) cutoff scale in a view point of four-dimensional effective theory. 
To search for the highest possible UV cutoff scale, we have evaluated the vacuum stability bound on the Higgs potential by solving renormalization group equation, whose details will be shown in Ref.~\cite{KNOOW:UED2013}.

\section{Higgs signal at the LHC in UED models}

\subsection{Prediction on Higgs signal in UED models}
The Higgs signal at the LHC can be divided into two parts, production and decay processes.
Higgs production at the LHC mainly comes from gluon fusion through the top loop.
Its production cross section is about ten times larger than other channels in SM case.
On the other hand, Higgs to diphoton and digluon decays are also induced as loop processes that are mainly constructed by top and W boson loops.
The branching ratio of the diphoton decay is ten times much smaller than the other decay channels, 
but we can see the diphoton signal clearly at the collider.
KK tower in the UED models affects such loop processes. 
Experimental data gives a constraints on signal strength that is defined by the ratio of the cross section,
\al{
   \frac{\sigma ^\text{exp}_{pp \rightarrow H \rightarrow X}}{\sigma ^\text{SM}_{pp \rightarrow H \rightarrow X}} ,
 }
where $X=\gamma \gamma,ZZ,WW,$ etc.
We compute it for the gluon fusion production channel in the UED models:
\al{
   \frac{\sigma ^\text{UED}_{gg\rightarrow H \rightarrow X}}{\sigma ^\text{SM}_{gg\rightarrow H \rightarrow X}} 
  &\simeq \frac{\Gamma ^\text{UED}_{H \rightarrow gg} \Gamma ^\text{UED}_{H\rightarrow X} /\Gamma ^\text{UED}_{H}}
          {\Gamma ^\text{SM}_{H \rightarrow gg} \Gamma ^\text{SM}_{H\rightarrow X} / \Gamma ^\text{SM}_{H}} \label{signal-st-2gamma},
}
where $\Gamma ^{UED/SM}_{H}$ is total decay width of the Higgs in UED/SM case and
\al{
  \label{Eq:GF}   \hat\sigma^\text{UED}_{gg\to H} &= \frac{\pi^2}{8 M_H} \Gamma ^\text{UED}_{H\rightarrow gg} \delta (\hat{s} - M_H^2), \\
  \label{Eq:Hgg}   \Gamma ^\text{UED}_{H\rightarrow gg}  &= K \frac{\alpha^2_s}{8\pi^2}\frac{M_H^3}{v^2_{EW}} |J^\text{SM}_t + J^\text{KK}_t|^2.
}
$K$ is K-factor which accounts for the higher order QCD corrections, $\alpha_s = \frac{g_s^2}{4\pi}$ is the fine structure constant for QCD,
 $v_{EW} \simeq$ 246 GeV is vacuum expectation value of Higgs in weak scale, and $J^\text{SM/KK}_t$ indicates the SM/KK top loop function, defined as in Ref.~\cite{Nishiwaki:2011gm, Nishiwaki:2011vi}.
The  diphoton decay width takes the following form:
\al{
\label{Eq:H2gamma}
\Gamma ^\text{UED}_{H\rightarrow \gamma \gamma} 
 &= \frac{\alpha^2 G^2_F M_H^3}{8\sqrt{2}\pi^3} \left| J^\text{SM}_W + J^\text{KK}_W + \frac43 (J^\text{SM}_t +J^\text{KK}_t) \right|^2,
}
where $\alpha = \frac{e^2}{4\pi}$ and $G_F$ are fine structure constant for QED and Fermi constant. 
The SM/KK $W$ boson loop function $J^\text{SM/KK}_W$ are also defined in Ref.~\cite{Nishiwaki:2011gm, Nishiwaki:2011vi}.
For the final states $X=ZZ/WW$, we can approximate as $\Gamma ^\text{UED}_{H\rightarrow ZZ/WW} \sim  \Gamma ^\text{SM}_{H\rightarrow ZZ/ WW}$ 
because Higgs decays into $ZZ/WW$ boson pair at the tree level, and hence KK loop contributions are negligible.
As an illustration, we show in FIG.1 the KK loop effect on the branching ratios and on the UED/SM ratio of the diphoton Higgs decay as well as of the digluon one, in the $ T^2/Z_2$ model.
The UED/SM ratio of $H \rightarrow gg$ (cyan line in left figure) is always enhanced in this range,
while that of $H \rightarrow \gamma \gamma$ (green line in left figure) is suppressed as already seen in Ref.~\cite{Nishiwaki:2011gm, Nishiwaki:2011vi}.
If we consider the KK compactification scale 1TeV, the signal strength of diboson decay is enhanced by a factor 1.5 from that of the SM, while the diphoton decay rate is suppressed by 10\%.
The signal strength $pp \to H \to \gamma\gamma$  becomes about 1.35 times larger than in the SM.
\begin{figure}[t]
\begin{center}
   \includegraphics[clip,width=170mm]{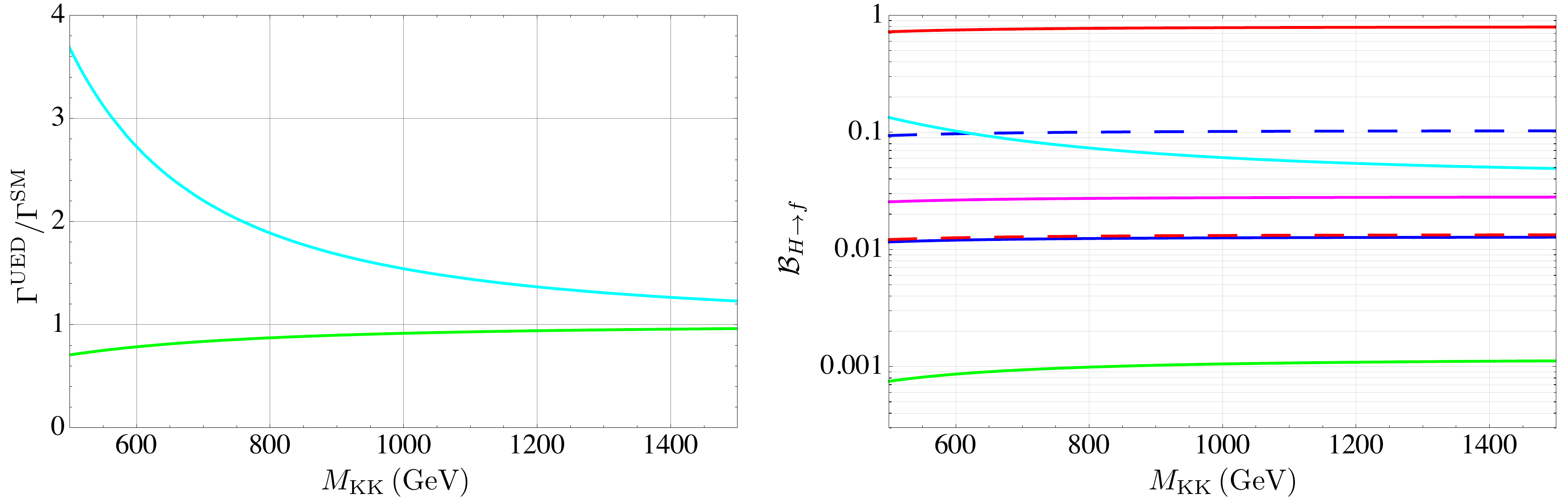}
\caption{UED/SM ratio of the Higgs decay rate (left) and the branching ratios (right) as functions of the KK scale $M_\text{KK}$ for the final states $bb$ (red), $cc$ (red dashed), $\tau\tau$ (magenta), $ZZ$ (blue), $WW$ (blue dashed), $gg$ (cyan), and $\gamma\gamma$ (green). Both left and right figures are for the $T^2/Z_2$ model. }
\label{Fig:UEDillustration}
\end{center}
\end{figure}
The branching ratios for the diphoton, diboson, 
and fermionic final states are suppressed compared with those in the SM, 
because of an enhacement of digluon decay rate, as shown in the right of FIG.1. 
The enhancement of $H \to gg$ rate is due to the KK top contributions in the loop diagrams.
The reason of the suppression in $H \rightarrow \gamma \gamma$ is as follows.
Each KK fermion mode is vectorlike, and hence has twice the degrees of freedom compared to its zero mode. 
Therefore their negative contributions to decay rate become larger than the positive ones coming from the KK $W$ loops.
The sum of all the KK loops gives negative contribution, while that of the SM ones is positive.

\subsection{Bound on KK scale from current data}
As shown above, the UED models give different production cross section in the gluon fusion (GF).
On the other hand, the other productions: the vector boson fusion (VBF),
the Higgs-strahlung (VH), and the associated production with a $t \bar{t}$ pair (ttH) are the same as in the SM.
The ATLAS and CMS have reported on the proportions of these production channels for each event categories of $H \rightarrow \gamma \gamma ,ZZ$ and $WW$~~\cite{ATLAS:2013gamma,ATLAS:2013Z,ATLAS:2013W,CMS:2013gamma,CMS:2013Z,CMS:2013W}.
We take these contributions into account. The details are given in~\cite{KNOOW:UED2013}.

\begin{figure}[t]
\begin{center}
   \includegraphics[clip,width=100mm]{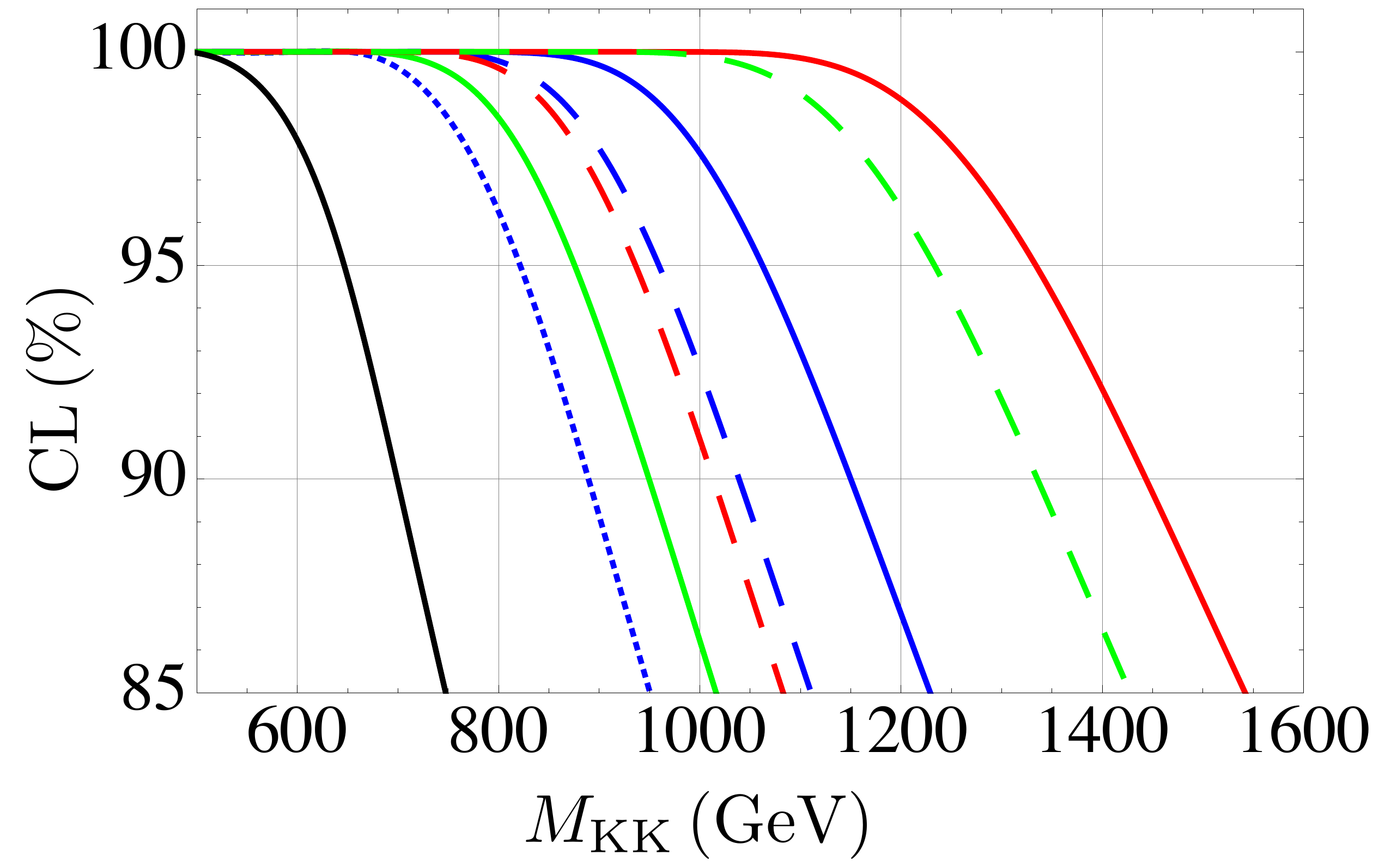}
\caption{Exclusion CLs of all the UED models as functions of the KK scale $M_\text{KK}$ by use of all the ATLAS and CMS results of $H \rightarrow \gamma \gamma,WW,ZZ$. 
The blue solid, dashed and dotted lines show the results in $T^2/Z_2, T^2/Z_2 \times Z_2'$ and $T^2/Z_4$ models; the red solid and dashed lines show those in $S^2$ and $S^2/Z_2$ models; 
the green solid and dashed lines show that in $RP^2$ and PS models; and the black line indicate that in mUED model.}
\label{Fig:Higgs}
\end{center}
\end{figure}

Figure \ref{Fig:Higgs} shows the bounds on the KK scale from all the ATLAS and CMS results of $H \to \gamma \gamma,WW,ZZ$ channels.
The blue solid, dashed, and dotted lines show the results in $T^2$-based models on $T^2/Z_2$, $T^2/(Z_2 \times Z_2')$, and $T^2/Z_4$, respectively. 
Similarly, the red solid and dashed ones in $S^2$-based models on $S^2$ and $S^2/Z_2$. 
The green solid and dashed ones on non-orientable manifolds $RP^2$ and $PS$. The black one for mUED model.
We can see from FIG.2 that the lower bound on the KK scale in the mUED model is around 650 GeV at the 95$\%$ confidence level (CL). 
The six dimensional models have the bounds that are heavier than in the mUED model.
This is because the number of state in each KK excited level in six-dimensional models is larger than in the mUED one. 
The bounds on the KK scale in the six-dimensional models $T^2/Z_2$, $T^2/(Z_2 \times Z_2')$ , $ T^2/Z_4$ are 1080 GeV, 950 GeV, and 830 GeV, respectively. 
Those on $S^2$ and $S^2/Z_2$ are 1330 GeV and 920 GeV, while on $RP^2$ and PS, 880 GeV and 1230 GeV, respectively.

\section{Electroweak precision constraint in UED models}

A measurement related to electroweak sector can be used to obtain indirect bounds on phenomenological models. 
The $S$ and $T$ parameters proposed by Peskin and Takeuchi~\cite{Peskin:1990zt,Peskin:1991sw} are very useful quantities for such a purpose. 
These parameters are represented as~\cite{Denner:1991kt}
\al{
\frac{\alpha S}{4 s_W^2 c_W^2} &=
		{\Pi^{\text{T}}_{ZZ}}'(0) + \frac{c_W^2 - s_W^2}{c_W s_W} {\Pi_{Z\gamma}^{\text{T}}}'(0)
		- {\Pi_{\gamma \gamma}^{\text{T}}}'(0), \\
\alpha T &= \frac{\Pi_{WW}^{\text{T}}(0)}{m_W^2} - \frac{\Pi_{ZZ}^{\text{T}}(0)}{m_Z^2}
		+ {2 c_W s_W} \frac{\Pi_{Z\gamma}^{\text{T}}(0)}{m_W^2},
} 
where $s_W$ and $c_W$ are sine and cosine of the weak mixing angle $\sin{\theta _W}$ and $\cos{\theta _W}$. 
The function ${\Pi^{\text{T}}_{ab}}^{(\prime)} (0)$ is the transverse component of the two point functions of the SM gauge bosons ($ab=WW, ZZ, Z\gamma, \gamma\gamma$), which is defined as 
\al{
\Pi^{\mu\nu}_{ab} (k) =  i\,\Pi^{\text{T}}_{ab} (k^2) \left( g^{\mu\nu} - \frac{k^{\mu}k^{\nu}}{k^2} \right) + i\,\Pi^{\text{L}}_{ab} (k^2) \frac{k^{\mu}k^{\nu}}{k^2} \label{gauge_twopointfunction},
}
where $k$ is the external momentum. ${\Pi^{\text{T}}_{ab}}' (k^2)$ is defined as $\frac{d}{dk^2} \Pi^{\text{T}}_{ab}(k^2)$. 
Several measurable quantities are represented as functions of the $S$ and $T$ parameters,
and from the global fit to the experimental results, the values of $S,T$ are estimated as~\cite{Baak:2012kk} 
\al{
S|^\text{exp}_{U=0} = 0.05 \pm 0.09,\quad
T|^\text{exp}_{U=0} = 0.08 \pm 0.07,
\label{ST_experimental}
}
with its correlation being $+0.91$, assuming that the $U$ parameter is zero. 
In an operator-analysis point of view, the $U$ parameter is represented as a coefficient of a higher dimensional operator involving the Higgs doublet than those for $S$ and $T$ in the UED models, 
and hence we ignore the effect in our analysis.

\begin{figure}[t]
\begin{center}
   \includegraphics[clip,width=100mm]{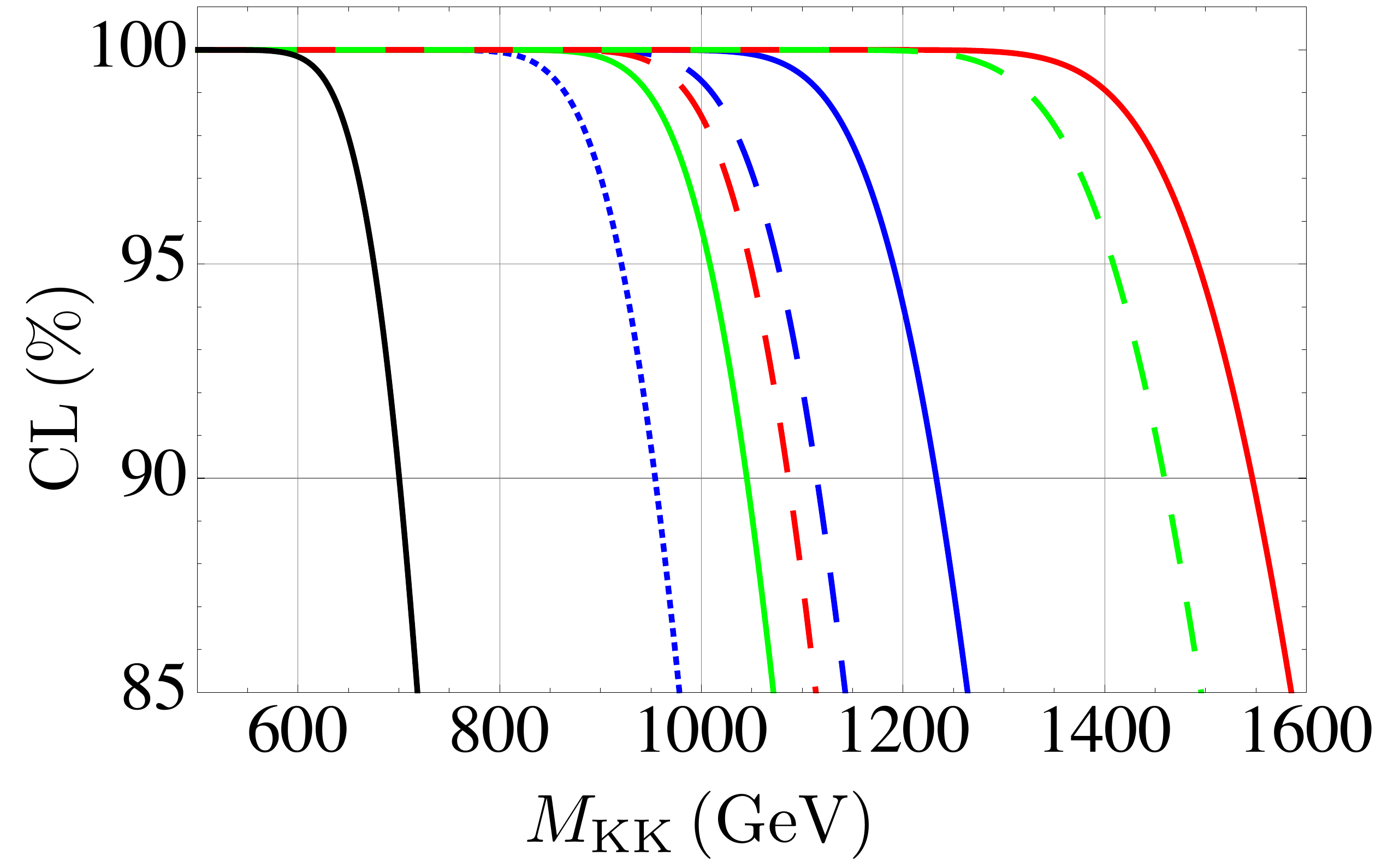}
\caption{Exclusion CLs of all UED models as functions of the KK scale $M_\text{KK}$ from the fit to the experimental results of $S$ and $T$ parameters. 
Colors denote the same as in Figure~\ref{Fig:Higgs}. }
\label{Fig:STresult}
\end{center}
\end{figure}
In the UED models, the forms of S and T are written as 
\al{
S = \sum_{s \atop \text{with }M_s < \Lambda} \left( S^{(\text{KK})}_{s,\text{boson}} + S^{(\text{KK})}_{s,\text{fermion}}  \right) + S_{\text{Higgs calibration}} + S_{\text{threshold}}, \label{Sform} \\
T = \sum_{s \atop \text{with }M_s < \Lambda} \left( T^{(\text{KK})}_{s,\text{boson}} + T^{(\text{KK})}_{s,\text{fermion}}  \right) + T_{\text{Higgs calibration}} + T_{\text{threshold}}, \label{Tform}
}
where the first two terms in Eqs.~(\ref{Sform}) and (\ref{Tform}) indicate the contributions of the KK particles via bosonic and fermionic loops, 
and the middle terms represent the effects from Higgs mass calibration, 
and the last terms show the threshold corrections via possible operators around the UV cutoff scale. 
In our analysis, in addition to the effects of the KK Higgs boson and the KK top quark~\cite{Appelquist:2002wb}, the effect of the KK gauge boson is newly taken into account. 
The detailed forms of each terms are found in Ref.~\cite{KNOOW:UED2013}.

In Fig.~\ref{Fig:STresult}, we show the bounds on the KK scales from the fit to the results in Eq.~(\ref{ST_experimental}). 
Colored lines correspond to each UED model in the same manner as shown in Fig.~\ref{Fig:Higgs}. 
We find that the lower bound on the KK scale in the mUED is around 700 GeV at the 95$\%$ CL. 
The bounds on the KK scale in $T^2/Z_2$, $T^2/(Z_2 \times Z_2')$, $ T^2/Z_4$, $S^2$, $S^2/Z_2$, $RP^2$ and PS are 1190 GeV, 1100 GeV, 900 GeV, 1500 GeV, 1050 GeV, 1020 GeV and 1410 GeV, respectively at the 95$\%$ CL.

\section{Summary}
We have estimated the two types of bounds on the KK scales in 5D and 6D UED models from the Higgs search at the LHC and from the electroweak precision data via the $S$ and $T$ parameters.
In the UED models, the contributions via loop diagrams including the KK top quarks and the KK gauge bosons modify the Higgs decay rate and production cross section. 
These contributions affect the Higgs signal strengths at the LHC. 
From the analysis on the results of Higgs signal strengths in the decay modes $H \to \gamma\gamma$, $H \to ZZ$ and $H \to WW$, 
we find that the lower bound on the KK scale in the mUED model is 650 GeV at the 95$\%$ CL, 
while those in the 6D models on $T^2/Z_2$, $T^2/(Z_2 \times Z_2')$, $ T^2/Z_4$, $S^2$, $S^2/Z_2$, $RP^2$ and PS are 1080 GeV, 950 GeV, 830 GeV, 1330 GeV, 920 GeV, 880 GeV and 1230 GeV, respectively. 
The KK excited states of the massive SM particles (top quark, Higgs boson and gauge boson) alter the $S$ and $T$ parameters. 
After evaluating the effects, the lower bound in the mUED turns out to be 650 GeV at the 95$\%$ CL, 
and the counterparts in $T^2/Z_2$, $T^2/(Z_2 \times Z_2')$, $T^2/Z_4$, $S^2$, $S^2/Z_2$, $RP^2$ and PS models are 1190 GeV, 1100 GeV, 900 GeV, 1500 GeV, 1050 GeV, 1020 GeV and 1410 GeV. 
Comparing the bounds from the Higgs signal search with those from the electroweak measurements, we find that the latter bounds are slightly severer than the former ones in the UED models for now.
However, in future the Higgs signal search at the LHC will put more strong constraints on the KK scales in the UED models.

\section{Acknowledgment}
We are grateful to Swarup Kumar Majee for discussions at the early stages of this work.
We thank Tomohiro Abe for useful comments on oblique corrections. 
K.N. is grateful for valuable discussions with Joydeep Chakrabortty
and Daisuke Harada, and for fruitful conversations with Anindya Datta
and Sreerup Raychaudhuri.
K.N. is partially supported by funding available from the Department
of Atomic Energy, Government of India for the Regional Centre for
Accelerator-based Particle Physics (RECAPP), Harish-Chandra Research
Institute.





\bibliographystyle{TitleAndArxiv}
\bibliography{referenceHPNP2013_2}

\end{document}